\documentstyle[prd,aps,preprint,psfig]{revtex}
\newcommand{\del}{\partial}

\newcommand{\K}{{\mathbf{k}}}

\newcommand{\f}{\frac}

\newcommand{\bb}{\bibitem}
\newcommand{\BF}{\begin{figure}\begin{center}}
\newcommand{\EF}{\end{center}\end{figure}}
\newcommand{\BE}{\begin{equation}}
\newcommand{\EE}{\end{equation}}
\newcommand{\BEA}{\begin{eqnarray}}
\newcommand{\EEA}{\end{eqnarray}}
\newcommand{\ti}{\textit}
\newcommand{\tb}{\textbf}
\begin{document}
%@twocolumnfalse\endcsname
%%
%%
\tighten
\draft
%%
%\flushright{YITP-00-01}
\title{How large is our universe?}		
\author{Kaiki Taro Inoue, Naoshi Sugiyama}
\address{National Astronomical Observatory
Division of Theoretical Astrophysics
2-21-1 Osawa, Mitaka, Tokyo 181-8588}
\date{\today}

\maketitle
\begin{abstract}
We reexamine constraints on the spatial size of closed toroidal 
models with cold dark matter and the cosmological 
constant from cosmic microwave background. We carry 
out Bayesian analyses using the Cosmic Background Explorer (COBE) 
data properly taking into account the 
statistically anisotropic correlation, i.e.,
off-diagonal elements in the covariance. We find that the COBE 
constraint becomes more stringent in comparison with that using only the
angular power spectrum, if the likelihood is marginalized over the
orientation of the observer. For some limited choices of orientations, the 
fit to the COBE data is considerably better than that of the infinite
counterpart. The best-fit matter normalization is 
increased because of large-angle suppression in the power and 
the global anisotropy of the temperature fluctuations.
We also study several deformed closed toroidal models in which 
the fundamental cell is described by a rectangular box.
In contrast to the cubic models, the large-angle power 
can be enhanced in comparison with the infinite counterparts if
the cell is sufficiently squashed in a certain direction. 
It turns out that constraints on some slightly deformed models are 
less stringent. 
We comment on how these results affect our understanding of 
the global topology of our universe.
\end{abstract}

%%%%%%%%%%%%%%%% INTRODUCTION %%%%%%%%%%%%%%%%%%%
\section{INTRODUCTION}
Is the universe finite or infinite?   If we assume the ``strong Copernican
principle'' (SCP) that the spatial hypersurface is globally and locally 
homogeneous and isotropic, then all three-spaces are classified by the sign 
of the curvature; spaces with a positive constant curvature are closed 
and finite in volume while spaces with a negative constant curvature 
or vanishing curvature are open and infinite. Hence, one might expect
that the sign of the spatial curvature completely determines the size of our universe. 

However, there is no particular theoretical reason for assuming
the global homogeneity and isotropy of the spatial hypersurface, since 
the Einstein equation itself cannot specify the global topology
of the space-time. In fact, there are a variety of 
cosmological models that satisfy
the ``weak Copernican principle'' for which 
the spatial section is locally homogeneous and isotropic but
not necessarily globally homogeneous and isotropic. These multiply
connected (MC) models (for a comprehensive review
 see \cite{LL95,InoueD,Levin01}) are topologically distinct but the 
metric is exactly the same as that of the 
simply connected counterparts with constant curvature. 
Therefore, these MC models are in harmony with the observational facts 
that support the Friedmann-Robertson-Walker (FRW) models, 
namely, the cosmic expansion, 
the relative amounts of primordial light elements, and the cosmic microwave
background. In order to answer whether or not  
the spatial section of our universe is finite, we need to determine 
not only the curvature, but also the global spatial topology of the
spatial hypersurface, since there are a number of closed compact 
spaces with a negative or a vanishing curvature. 

If the spatial section is sufficiently small, then all the  
fluctuations beyond the comoving size of the space  
are suppressed.  It has been claimed by several authors 
that the ``small universes''\cite{Ellis86} in which the 
topological identification scale is equivalent to or smaller 
than the horizon scale have been observationally ruled out
because large-angle power of the cosmic microwave background 
(CMB) temperature fluctuations
is significantly suppressed \cite{Sokolov93,Staro93,Stevens93,Oliveira95a}
and does not fit to the Cosmic Microwave Background Explorer (COBE)
Differential Microwave Radiometer (DMR) data. 

However, for low matter density models $\Omega_0<1$,  
constraints on the size of the spatial section 
are less stringent, since the bulk of large-angle temperature fluctuations
can be produced in the curvature or $\Lambda$ dominant era,
leading to less stringent suppression in the large-angle power
\cite{Inoue99,Aurich99,Inoue00a,CS00}.

On the other hand, it has been pointed out that 
constraining closed MC models 
using only the angular power spectrum is not sufficient, 
since it does not contain information about the anisotropic component of 
statistically averaged fluctuations \cite{Bond00b}. 
In fact, temperature fluctuations in the sky for the MC models 
(except for the projective three-space $R P^3$) are described by an anisotropic 
Gaussian random field for a particular choice of position and
orientation of the observer, provided that each Fourier mode is
independent and obeys Gaussian statistics as predicted by the standard
inflationary scenario. Therefore, in order to
constrain the MC models, one must compare all
anisotropic and inhomogeneous 
fluctuation patterns for every possible choice of 
orientation and position of the observer with the
observed data, which needs time-consuming Monte Carlo simulations. 
Based on the study of several closed
hyperbolic (CH) models,  \cite{Bond00b} claimed that 
all ``small universes'' are ruled out 
because the anisotropic correlation 
patterns are at variance with the COBE DMR data. 

Nevertheless, for CH models, a subsequent analysis confirmed that 
the fit to the data depends not only on the orientation  
but also on the position of the observer. In some limited places,
the likelihood is considerably higher than that 
of the simply connected counterpart\cite{Inoue01b}. 
Even if the place at which the pixel-pixel two-point 
correlation agrees with the data is limited, the model cannot be ruled
out if the fit to the data is sufficiently good at certain places.

On the observational side, there has been mounting    
evidence that the spatial section of our universe is almost flat
\cite{Perl99,Riess99,Melch00,Jaffe01}. 
If the background space is exactly flat, the topology 
of the orientable closed space is limited to six kinds \cite{LL95}. 
The simplest space is the toroidal space $T^3$. Although 
closed flat MC models need a ``fine-tuning'' 
that the present horizon scale is comparable to
the topological identification scale $L_t, 
$\footnote{In this paper, we define $L_t$ as the comoving length 
of a closed geodesic which connects a point $x$ and another 
point $g x$ in the universal covering space where $g$ is a 
generator of the discrete isometry group $\Gamma$ that defines
a Dirichlet fundamental domain (i.e., a fundamental cell).} it is of crucial 
importance to give a lower bound on the 
size of the spatial section of these models, since we have 
no prior knowledge of the spatial topology of the universe.

In this paper, we improve the previous COBE bounds on the size of the spatial
hypersurface of closed toroidal flat cold dark matter models with a
consmological constant ($\Lambda$CDM)
\cite{Inoue01a}by carrying out Bayesian analysis properly taking account of
statistically anisotropic correlation that cannot be described by the
angular power spectrum $C_l$. 
The temperature fluctuations are calculated by solving the linear evolution of 
the photon-baryon fluid from the 
COBE scale to the acoustic oscillation scale. We estimate the 
best-fit normalization of the power of the matter density fluctuations at 
the $8\textrm{h}^{-1}$Mpc scale $\sigma_8$ in comparison 
with those of simply connected $\Lambda$CDM 
models. Throughout this paper we assume that 
$\Omega_m\!=\!0.3,\Omega_\Lambda\!=\!0.7$.  In Sec.II, we give a brief
account of the temperature anisotropy on large to 
intermediate scales and we study how the size and the shape of the 
fundamental cell affect the large-angle anisotropy. 
In Sec.III, we carry out full Bayesian analyses using the COBE DMR data
based on pixel-pixel correlation and obtain the
likelihoods and the best-fit normalizations $\sigma_8$. 
In Sec.IV, we discuss the viability of other closed flat models, 
almost-flat closed spherical and almost-flat closed hyperbolic models.  
In Sec.V, we draw our conclusions and discuss the possibility of
detecting the signature of the finiteness of the spatial geometry of the
universe in the future.  
\section{Temperature Anisotropy}
In the Newtonian gauge, the $l$th multipole of the 
primary temperature anisotropy in the FRW models projected onto the
sky at the conformal time $\eta$ is written in terms of the monopole $\Theta_0$,
and the dipole $\Theta_1$ of temperature fluctuation and the Newtonian
potential $\Psi$ at the last scattering and the time evolution of 
$\Psi$ and the Newtonian curvature $\Phi$\cite{HSS}, 
\BEA
\f{\Theta_l(\eta,k)}{2l+1}&=& [\Theta_0+\Psi](\eta_*,k)
        j_l(k (\chi(\eta)-\chi(\eta_*))) + \Theta_1(\eta_*,k)
        \f{1}{k}\f{d}{d \eta}j_l(k(\chi(\eta)-\chi(\eta_*))) \nonumber\\
        & + & \int_{\eta_*}^{\eta} \Bigl (\f{\del \Psi}{ \del \eta'}  -
        \f{\del \Phi}{\del \eta'} \Bigr)
        j_l(k(\chi(\eta)-\chi(\eta')))
        d\eta' , \label{eq:anisotropy}
\EEA
where $k$ denotes the wave number, $\eta_*$ is the last scattering
conformal time, the radial distance is $\chi(\eta)=\eta$ and $j_l$ represents the 
$l$th-order spherical Bessel function.
The first term in the right-hand side in 
equation (\ref{eq:anisotropy}) represents the
``ordinary'' Sachs-Wolfe (OSW) effect caused by the
fluctuations in the gravitational potential at the last scattering.  
The second term describes the Doppler shift owing to the 
bulk velocity of the photon-baryon fluid. Note that this term 
is negligible at superhorizon scales $k \!\eta\!<1$ since radiation pressure
cannot play any role and cosmic expansion damps away the initial 
velocity. The third term represents the integrated Sachs-Wolfe (ISW) effect
caused by the decay of the gravitational 
potential.

In order to constrain the FRW models that yield 
isotropic Gaussian fluctuations, one needs only two-point 
correlations of temperature fluctuations $\Delta T/T$ 
in the sky, which can be characterized 
by the angular power spectrum $C_l=\langle |a_{lm}|^2 \rangle$, 
\BE
\f{2 l + 1}{4\pi} C_l \propto
\f{1}{2 \pi^2} \int_0^\infty {d k \over k}
 k^3 \f{|\Theta_l(\eta_0,k)|^2 }{2l+1}, \label{eq:cl} 
\EE
where $a_{lm}$ represents the expansion coefficient with respect to
the spherical harmonic $Y_{lm}$, 
$\langle \rangle$ denotes an ensemble average over the initial 
fluctuations, and $\eta_0$ is the present conformal time. 
Note that $C_l$ is invariant under any $SO(3)$ transformations. 

Now, let us consider a closed toroidal model $T^3$ in which 
the spatial hypersurface is described by a cube
with side $L$ in the Euclidean three-space whose 
opposite faces are identified by translations. 
Then the wave numbers of the square-integrable 
eigenmodes of the Laplace-Beltrami operator are restricted to
the discrete values $k_i=2 \pi n_i/L$, ($i=1,2,3$) where $n_i$'s run
over all integers. Equation $(\ref{eq:cl})$ now reads
\BE
\f{2 l + 1}{4\pi} C_l  \propto
\f{1}{L^3} \sum_{\K\ne 0} \f{|\Theta_l(\eta_0,k)|^2 }{2l+1}.
\label{eq:cl2} 
\EE
Note that all the modes whose wavelength 
is longer than $L$ are not 
where $a_{lm}$ represents the expansioallowed (``mode cutoff'').
From now on, we restrict ourselves to the adiabatic scale-invariant 
Harrison-Zel'dovich spectrum $n\!=\!1$ [i.e., ${\cal{P}}_\Phi(k)
\!=\!k^3\langle |\Phi(0,k)|^2 \rangle\!=\!const.$]
as predicted by the standard inflationary scenario. Note that $\Delta
T_l=[l(l+1)C_l/(2\pi)]^{1/2}$ becomes almost flat for the
Einstein-de Sitter universe. For low-density FRW models, the
large-angle power is boosted owing to the ISW effect, since $\Lambda$
dominant epochs come earlier. However,  
for $T^3$ models, the OSW contribution to 
the large-angle power is suppressed because 
all modes whose wavelength is longer than the cutoff scale
$L$ are not allowed. The angular scale below which the power is
suppressed is approximately given by $l_{cut}=2 \pi \eta_0/L-1$
\cite{Inoue01a}.
Interestingly, for low-density $T^3$ models with moderate size of the
cell, the excess 
power owing to the ISW contribution is almost canceled out by
the suppression owing to the mode cutoff. 
As shown in figure ({\ref{fig:1}}), the suppression in 
the large-angle power is not prominent even for a model with 
$\epsilon\equiv L/2 \eta_0=0.2$. 

\BF
\centerline{\psfig{figure=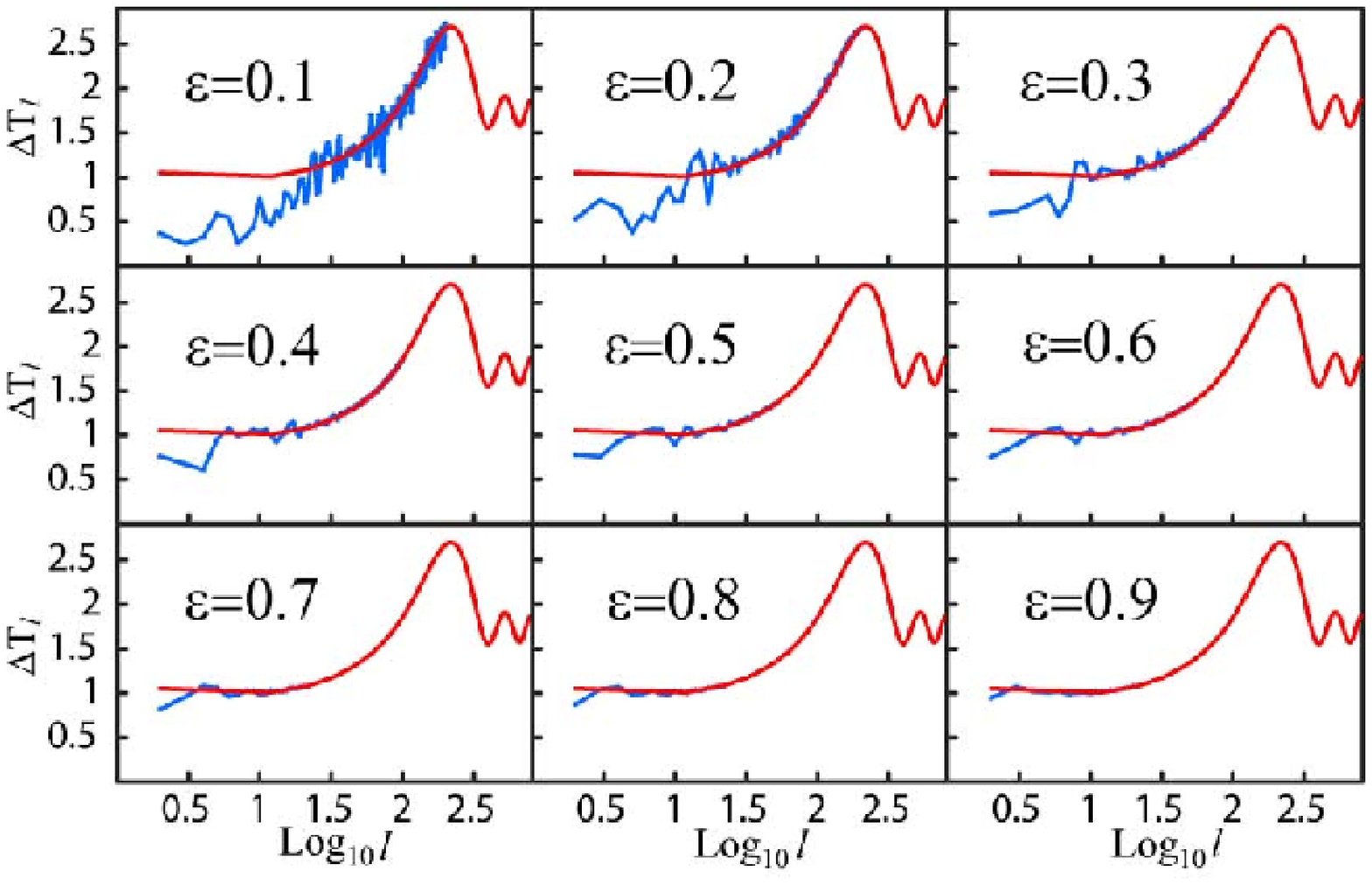,width=17cm}}
\caption{Plots of the angular power spectrum $\Delta T_l=
[l(l+1)C_l/(2\pi)]^{1/2}$  for cubic toroidal
models with various linear sizes parametrized by the ratio of the length
of the side to the diameter of the last scattering surface in the
comoving coordinates:$\epsilon=L/2 \eta_0$ 
in comparison with that for 
the simply connected counterpart (solid curve).  $\Delta T_l$
is normalized by $\Delta T_{10}$ for the simply connected 
models. Cosmological parameters are 
$\Omega_\Lambda=0.7$, $\Omega_{CDM}=0.26$, 
$\Omega_b=0.04$, and $h=0.7$. }
\label{fig:1}
\EF

\BF
\centerline{\psfig{figure=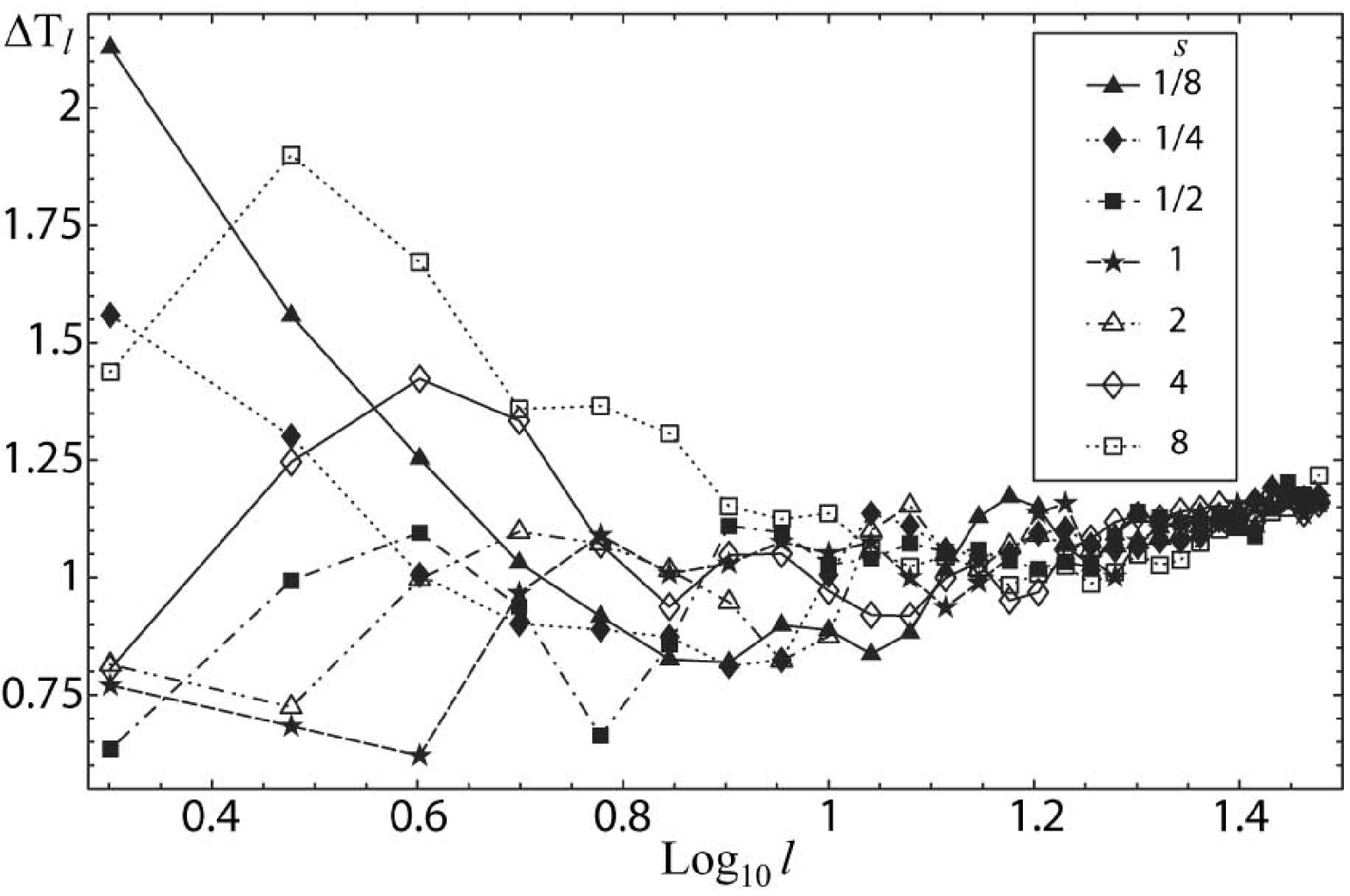,width=15cm}}
\caption{Plots of the angular power spectrum $\Delta T_l=
[(l+1)C_l/(2\pi)]^{1/2}$  for deformed toroidal
models whose fundamental cell is described by a rectangular box
with sides $L_1\!=\!L_2,L_3$. The volume is fixed to $V=(0.4\times 2 \eta_0)^3$.
The degree of deformation is parametrized 
by $s\equiv(L_1=L_2)/L_3$. 
$\Delta T_l$ is normalized by $\Delta T_{10}$ for the simply connected 
counterpart.  In the small-angle limit $l \gg 1$, the angular power spectrum
converges to that of the simply connected counterpart. Cosmological 
parameters are the same as in figure 1.}

\label{fig:2}
\EF

Next, we consider models in which the fundamental cell 
is described by a rectangular box $D$ with length $L_i$ of sides. We
call these models the ``pancake' type or the ``cigar'' type if
$D$ has a relatively long side in one direction, or long sides in two orthogonal 
directions, respectively. As shown in figure 2, 
the large-angle power is enhanced for models with a sufficiently squashed  
cell. It seems that the enhancement is rather surprising since Fourier modes whose
wavelength is larger than the size of the cell are not allowed. 
The apparent enhancement is related to the number density 
of the Fourier modes. In order to see this, let us
consider the number
function defined as the number of eigenmodes of the Laplace-Beltrami
operator whose eigenvalue is less than $E=k^2$, which is described by Weyl's
formula in the small-scale limit
$N(E)=V k^3 /2 \pi^2$, where $V$ denotes the volume of the three-space. 
For globally ``almost'' isotropic spaces like cubic toroidal spaces, the 
Weyl's formula gives a good approximation of $N(k)$ in the large-scale
limit as well. However, for globally ``very'' anisotropic 
spaces \footnote{The degree of global anisotropy in the geometry 
can be measured by the ratio 
of the maximum geodesic shortest distance between two arbitrary points 
to the characteristic length scale $V^{1/3}$. }, 
the deviation from the Weyl formula
is prominent in the large scale limit. 
For instance, the number function is approximately given by $N\propto k$
for the ``cigar'' type and $N\propto k^2$ for the 
``pancake'' type \cite{Inoue00b}. The discrete wave number is given by 
\BE
k=2\pi \sqrt{\f{n_1^2}{L_1^2}
+\f{n_2^2}{L_2^2}
+ \f{n_3^2}{L_3^2}
},
\EE
where each $n_i$ runs over all integers. Hence, for $L_2,L_3\ll L_1$,
the modes on large scales $k \ll 1/L_2,1/L_3$ are given by $k=2 \pi
n_1/L_1$, leading to $N(k)\sim L_1 k /\pi$. Similarly, if
$L_3 \ll L_1 \sim L_2$, the modes on large scales $k \ll L_3$ are
given by $k=2 \pi \sqrt{\f{n_1^2}{L_1^2}
+\f{n_2^2}{L_2^2}}$, then  $N(k)\sim L_1 L_2 k^2/4 \pi$. 
Because the power is 
normalized by the total volume, the contribution 
of modes on large scales are relatively boosted, 
resulting in an enhancement in the large-angle power. 
Interestingly, for slightly squashed models, the reduction in 
the large-angle power owing to the mode cutoff is almost canceled
out by the enhancement owing to the squeezing of the space. 

Although we have assumed that the
primordial power spectrum is scale-invariant even in the case of
deformed models, the non-trivial boundary condition on the 
mode functions of inflaton fields may cause 
a deviation from the scale-invariant spectrum. However, 
such a deviation is expected to be small, since the background
space time has a high degree of symmetry locally
at the inflationary epoch that is relevant to the 
$n=1$ spectrum.

So far, we have studied the effect of the nontrivial global topology on
the large-angle power. However, the angular power spectrum 
is not sufficient for constraining the models 
since the temperature fluctuations in the MC models are 
not statistically isotropic in general; the effect of 
off-diagonal elements $\langle a_{lm} a^*_{l'm'}\rangle$ for $l\ne l'$ 
or $m\ne m'$ which are not SO$(3)$ invariants cannot be negligible.
Therefore, it is necessary to perform Bayesian analyses taking account of 
these off-diagonal elements for each possible orientation of the
observer, since the toroidal models are globally anisotropic. 
\section{Bayesian analysis}
Any Gaussian temperature fluctuations in the sky can be 
characterized by a covariance matrix whose elements 
at pixel $i$ and pixel $j$ are,
\BE
M_{ij}=\langle T_i T_j \rangle =
\sum_{l l' m m'} \langle a_{lm}a^*_{l'm'} 
\rangle W_l W_{l'}Y_{l m}(\hat{\textbf{n}}_i)
Y_{l' m'}(\hat{\textbf{n}}_j)+\langle N_i N_j \rangle
\label{eq:M}
\EE
where $a_{lm}$ is an expansion coefficient with respect to a
spherical harmonic $Y_{lm}$, and $\langle \rangle$ denotes an ensemble
average taken over all initial fluctuations for a fixed 
orientation of the observer. 
$T_i$ represents the temperature at pixel $i$, $W_l^2$  is the
experimental window function that includes the effect of 
beam smoothing and finite pixel size,
$\hat{\textbf{n}}_i$ denotes the unit vector toward the center of pixel $i$,
and $\langle N_i N_j \rangle$ represents the noise covariance between 
pixel $i$ and pixel $j$. 

If we assume a uniform prior distribution for the cosmological
parameters, from Bayes' theorem, the probability distribution 
function of a set of $C^{l' m'}_{l m}\equiv \langle a_{lm}a^*_{l'm'}
\rangle$ for a random Gaussian field (not necessarily isotropic) is
given by
\BE
\Lambda(C^{l' m'}_{l m}|\vec{T})\propto\f{1}{\det^{1/2}M(C^{l' m'}_{l m}) } 
\exp \Biggl (\f{1}{2}\vec{T}^T\cdot M^{-1}(C^{l' m'}_{l m}) \cdot \vec{T}\Biggr). 
\label{eq:LL}
\EE

In the following analysis, we use the inverse-noise-variance-weighted
average map of the 53A, 53B, 90A, and 90B  
COBE DMR channels. To remove the emission from the galactic
plane, we use the 
extended galactic cut (in galactic coordinates) \cite{Banday}.
After the galactic cut, the best-fit monopole and dipole are removed  
using the least-squares method.
To achieve efficient analysis in computation, we compress 
the data at ``resolution 6'' $(2.6^o)^2$ ($l\sim 72$)  
pixels into one at ``resolution 4'' $(10.4^o)^2$  ($l\sim 18$) 
pixels for which there are 384 pixels in the
celestial sphere and 248 pixels surviving the extended galactic cut.
The window function is given by $W_l=G_l F_l$ where $F_l$ are the Legendre
coefficients for the DMR beam pattern \cite{Lineweaver} and $G_l$ 
are the Legendre coefficients for a circular 
top-hat function with area equal to the
pixel area, which account for the pixel smoothing effect \cite{Hinshaw}.
We also assume that the noise in the pixels is 
uncorrelated from pixel to pixel, 
which is known to be a good approximation\cite{Tegmark-Bunn}.
\BF
\centerline{\psfig{figure=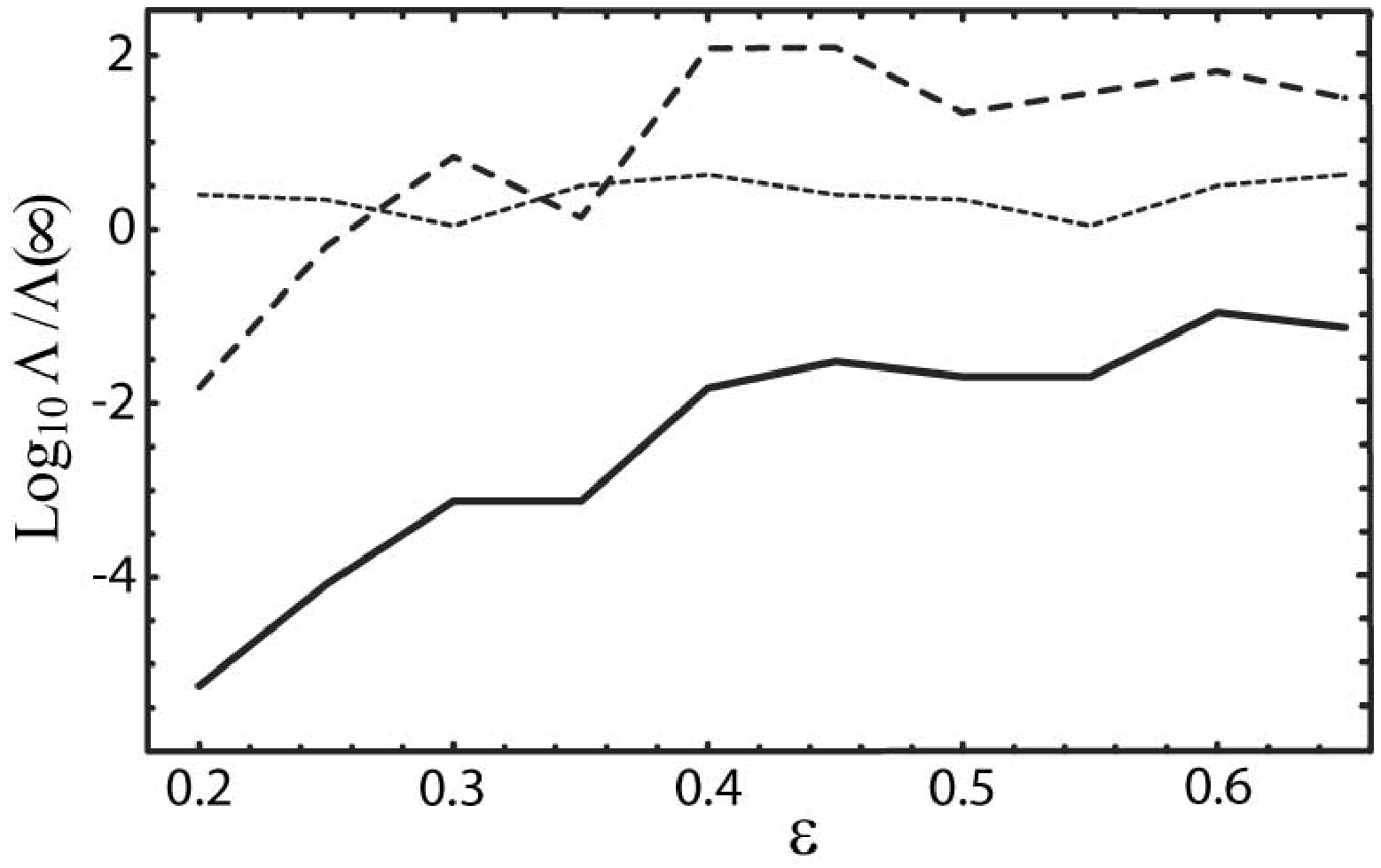,width=11cm}}
\caption{COBE constraints on cubic models. 
The solid curve represents the likelihoods $\Lambda$   
marginalized over 24000 orientations of the observer 
and the normalization for various relative linear size $\epsilon=L/2 \eta_0$.
The dashed line shows the maximum values.  The dotted line represents the likelihoods 
computed by neglecting the anisotropic components.  All the likelihoods are
normalized by that of the simply connected counterpart
$\Lambda(\infty)$.
 Cosmological parameters are the same as in figure 1. }
\label{fig:3}
\EF

Because the toroidal $T^3$ model is globally anisotropic, the two-point
pixel-pixel correlations depend on the orientation of the coordinate axes. 
Therefore, we need to compute the likelihood 
function for each orientation $\alpha$ of the
observer. Assuming a constant distribution for $\alpha$ and 
the normalization $\sigma$, the likelihood is given by $\Lambda=\int
\Lambda(\alpha,\sigma)d\alpha d\sigma $, where $d \alpha$ denotes the
volume element of a Lie group SO$(3)$ with Haar measure.

As shown in figure 3, the angular power spectrum alone does not
give a stringent constraint $\epsilon>0.3$ on the size of the cell. 
However, if we use a likelihood function taking account of anisotropic
two-point correlations marginalized over the 
orientation of the observer, we obtain a rather stringent constraint.

Because there is no upper limit for $\epsilon$, we choose a
variable $N=\pi/6\epsilon^3= $(volume of the observable region 
at present)/(volume of space at present) 
as a physical quantity that is assumed to be constantly distributed.
We find that the logarithm of relative likelihoods
$\log_{10}[\Lambda(N)/\Lambda(N=0)]$ for cubic models ($0.2 \le \epsilon
\le 0.65$) can be well fitted by a polynomial function $-b N^a$ where $a\!=\!
0.46, b\!=\!0.78$ which are obtained by minimizing $\chi^2$.
By marginalizing over $N$, we find a constraint $N\!\lesssim\!2.4$ ($\epsilon\!>\!0.6$)
at 68\% CL, and $N\!\lesssim\!8.2$ ($\epsilon\!>\!0.4$) at 95\% CL 
corresponding to $\log_{10}\Lambda/\Lambda(\infty)\sim -0.9$,
$\log_{10}\Lambda/\Lambda(\infty)\sim -2.0$, respectively.
The constraint on the size of the spatial section 
is more stringent compared with the previous analysis \cite{Inoue01a}
in which fluctuations on smaller angular scales $l\!\ge\! 10$ have not
been taken into account. 

It should be emphasized that 
the likelihoods depend sensitively on the 
orientation of the observer, especially for models with small
sides $\epsilon \ll 1$. The maximum value of the likelihood far
exceeds the mean by a factor $\Lambda_{\max}/\Lambda=
10^{2.5-3.9}$ for $0.2 \le 
\epsilon \le 0.7$ which suggests that almost all the orientations 
are ruled out. For instance,
$\Lambda_{\max}/\Lambda\sim10^{3.9}$ implies that the likelihood
attains high values only when the axes are oriented with a
precision of $\sim 5^o$ which is 
equal to one half the size of one pixel.\footnote{For 24000 realizatons of
orientation, the distance between a pair of the nearest two points 
on SO$(3)$ is approximately a cubic root of 
[volume of SO$(3)$]/24000/(number of symmetries of $T^3$)
= $(8 \pi^2/24000/24)^{1/3}\times 180/\pi\sim 3^o$. Therefore, each peak 
($\sim\!5^o$) is represented by just several points. 
We have checked the convergency of the
calculations by increasing the number of points near the peaks
to several hundreds, which resulted in just 10 percent 
difference in the relative likelihoods. }

To see the orientations that give the enormous likelihood,
we have plotted the orientations in the skymap 
that give a large likelihood. Except for 
$\epsilon=0.2$, in which a weak correlation to the galactic pole is found,
no particular preferred orientation is found (figure 4).
\pagebreak
\BF
\centerline{\psfig{figure=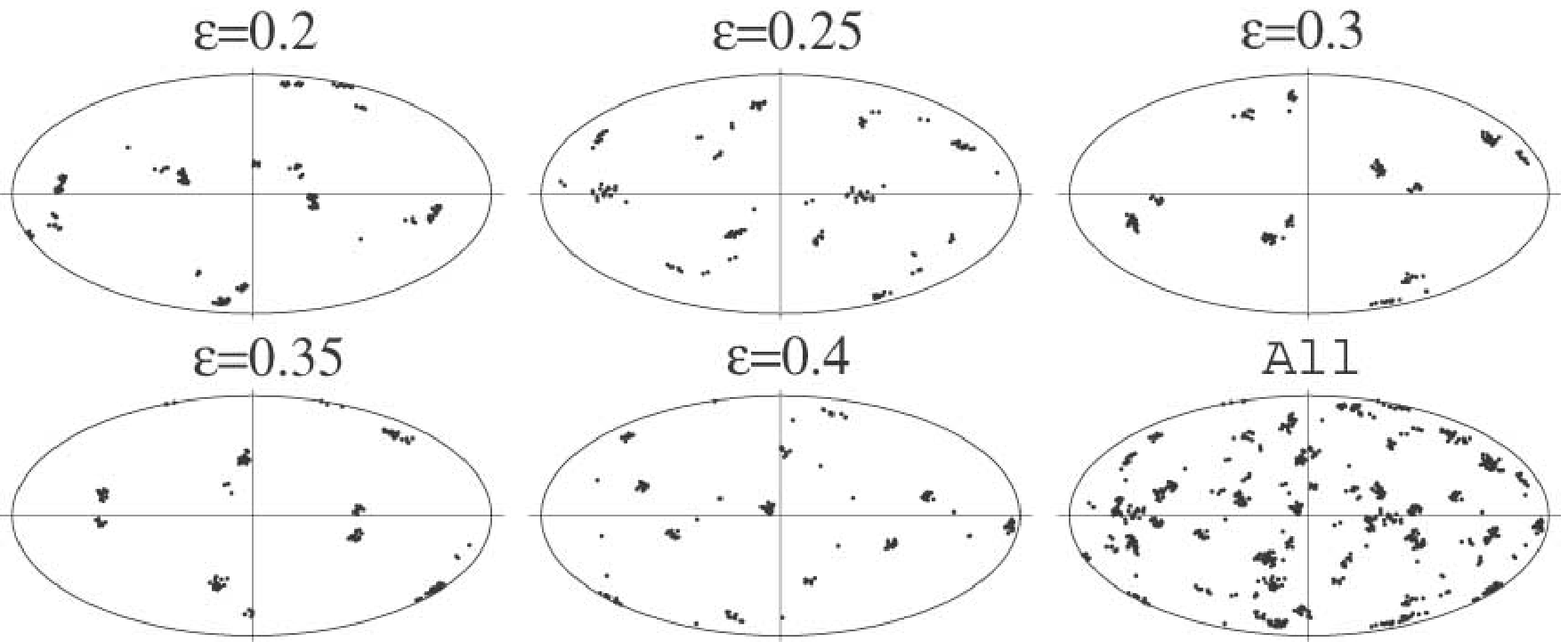,width=15cm}}
\caption{Best-fit orientations in the skymap 
for five cubic models.  Each orientation is 
described by a set of six points that correspond to directions to the 
centers of the face of the cube that are preserved by the symmetries of
$T^3$. A total of 40 orientations
that give more than 80 percent contribution to the 
averaged likelihoods are plotted for each model.
All the orientations for the five 
models are shown together in the last figure. 
The Aitoff projection is used. The center of the oval 
represents the galactic center.}
\label{fig:4}
\EF

\BF
\centerline{\psfig{figure=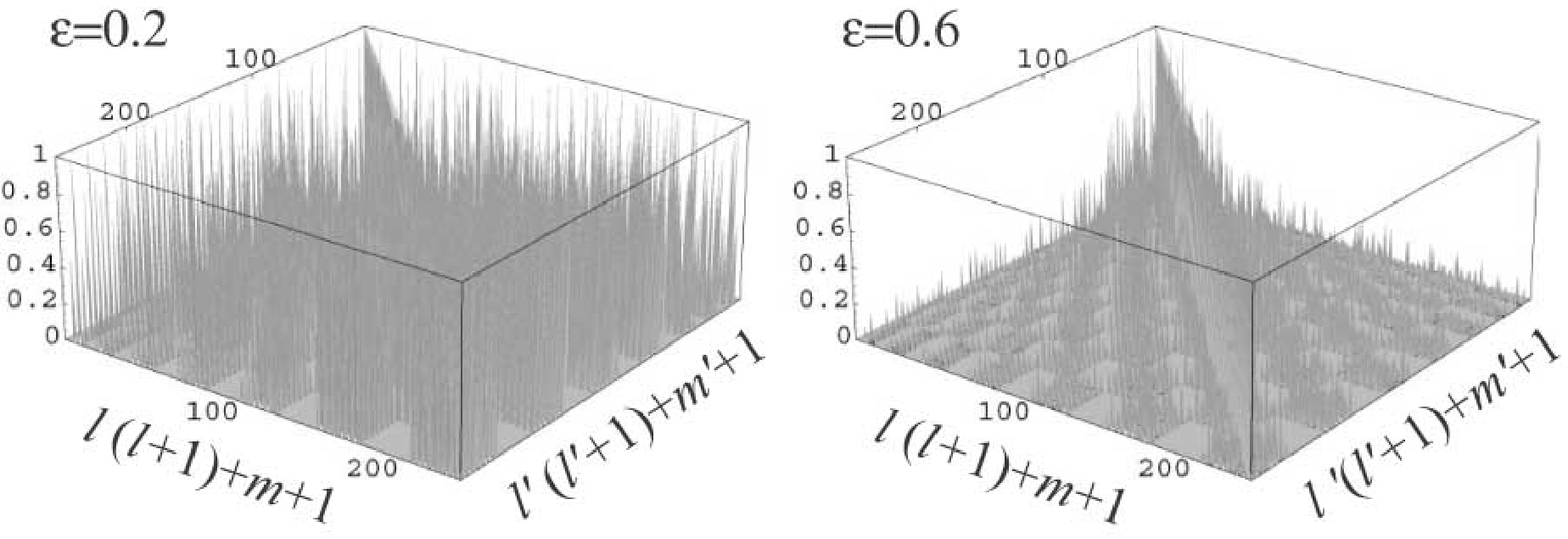,width=17cm}}
\caption{The contribution of off-diagonal elements for cubic models. 
The plotted lines represent $|\langle a_{l m}a^*_{l' m'}\rangle|
/\sqrt{\langle a_{l m}a^*_{l' m'} \rangle \langle a_{l m}a^*_{l'm'}\rangle}$ 
for $2\!\le\! l \!\le\! 15$, in ascending order of 
$l(l+1)+m+1$ and $l'(l'+1)+m'+1$ for two types of cubic models.}
\label{fig:5}
\EF
\pagebreak

In the previous literature, the effect of the global anisotropy has been
considered to be small for cubic cases, because the contribution of 
off-diagonal elements is quite small. This argument is valid as long as one 
considers the averaged values 
over all the possible set of parameters $(l,l',m,m')$. Let us consider
a set of orthogonal coordinates $(x,y,z)$ in which all the faces of the cube
are perpendicular to one of the coordinate axes. Because of the 
symmetries of $T^3$, $\langle a_{l,m}a^*_{l',m'} \rangle$ vanishes
if $l+l'$ is even and $(m+m')/2$ is even. In fact, the
contribution of off-diagonal elements relative to the diagonal elements 
are small as 0.042 to 0.015 for $\epsilon\!=\!$0.2 to 0.6 for 
an ensemble of all values of $(l,l',m,m')$. However, as shown in figure
5, each nonzero component of off-diagonal elements is not so small
(0.33 to 0.12 for $\epsilon\!=\!$ 0.2 to 0.6). Although any rotation 
of the coordinates gives a different structure in the off-diagonal
components, the spiky structure cannot be completely  
vanished by any rotation. 

One might argue that computing a likelihood marginalized over 
every orientation is meaningless and the constraint should be given by 
the maximum value of the likelihood instead of the marginalized value. 
Then the constraint becomes less stringent $\epsilon>0.25$.
However, if we assume that the universe does not prefer a specific 
orientation, the cases of the ``failed orientations'' should be taken 
into account. By deforming the shape of the fundamental cell,
or by changing the amplitude of the fluctuations, 
there might be a chance that the fluctuation fits the 
data almost perfectly, since the data themselves are anisotropic.
In order to check this, a similar analysis has been 
carried out for deformed toroidal models whose
fundamental cell is described 
by a rectangular box with sides $L_1=L_2$ and $L_3$. 
As shown in figure 6, the ``pancake'' type models $(L_1=L_2)/L_3>1$ 
give much better fits to the data. For instance, for $s=10^{1/2}$, 
the likelihood is improved by 10 times in comparison with
the cubic model with the same volume $V=(0.4\times 2 \eta_0)^3$. The corresponding 
number of copies of the fundamental cell inside the last 
scattering surface in the comoving coordinates is $N\!=\!8$ and 
the minimum linear size of the cell is equal to $L_3/2 \eta_0=0.19$. 
\pagebreak
\BF
\centerline{\psfig{figure=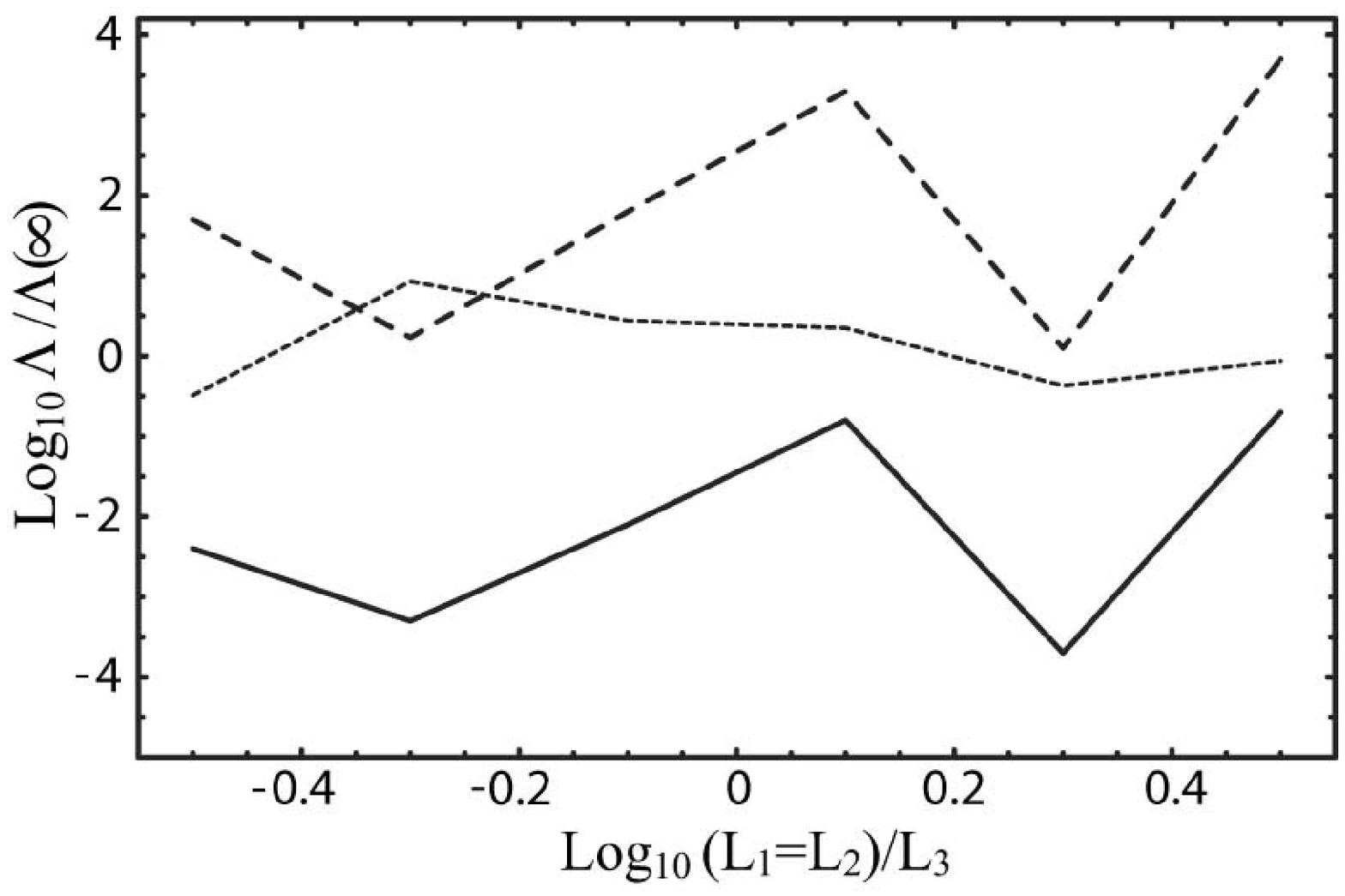,width=11cm}}
\caption{COBE constraints on deformed models
whose fundamental cell is described by a rectangular box with sides
$L_1\!=\!L_2$ and $L_3$. The volume is fixed to $V=(0.4\times 2 \eta_0)^3$. 
The solid curve represents the likelihoods $\Lambda$   
marginalized over 24000 orientations of the observer 
and the normalization for various models. 
The dashed line shows the maximum values.  The dotted line represents the likelihoods 
computed by neglecting the anisotropic components. 
All the likelihoods are normalized by that of the simply connected counterpart
$\Lambda(\infty)$. Cosmological parameters are the same as in figure 1. }
\label{fig:6}
\EF

In the previous literature, it has also been 
argued that an increase in the best-fit 
normalization constant is related to
the large-angle suppression owing to the finite size of the spatial
section \cite{CS00}. From our analyses, it is found that   
inclusion of off-diagonal term can also 
give a comparable increase in the best-fit normalization constant as well.
For instance, the best-fit normalization $\sigma_8$ is increased by
$\sim10$ percent for $\epsilon=0.6$ (figure 7).
The effect is relevant to the globally anisotropic geometry of the
background space (see the appendix) which has 
been found for closed hyperbolic models (although the effect of the
global inhomogeneity is neglected) \cite{Bond00b}.

If the best-fit normalization is too large, 
the choice of orientation is less probable (figure 8) (see the
appendix). We have checked that 
the anticorrelation between the likelihood and the best-fit
normalization becomes weaker and the plotted points get more aligned
as the comoving size of the spatial section increases. 

\BF
\centerline{\psfig{figure=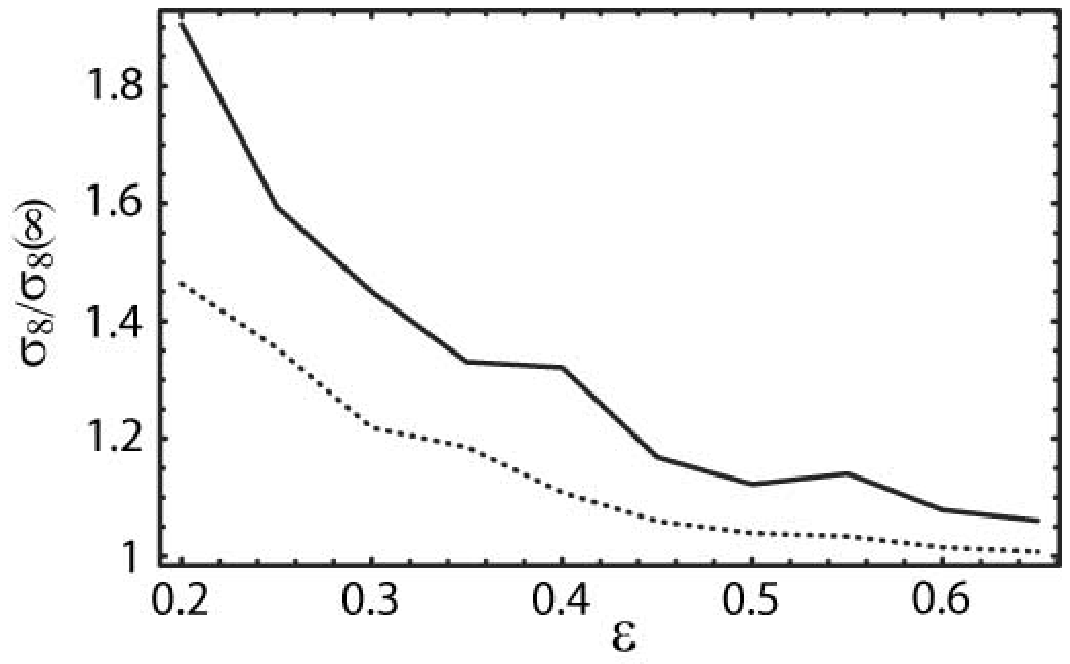,width=10cm}}
\caption{Plots of the ratio of the best-fit normalization of the cubic
models to that of the simply connected counterpart (solid curve) and those
obtained by an analysis in which the covariance matrix is computed by 
neglecting the off-diagonal terms (dotted curve). }
\label{fig:7}
\EF

\BF
\centerline{\psfig{figure=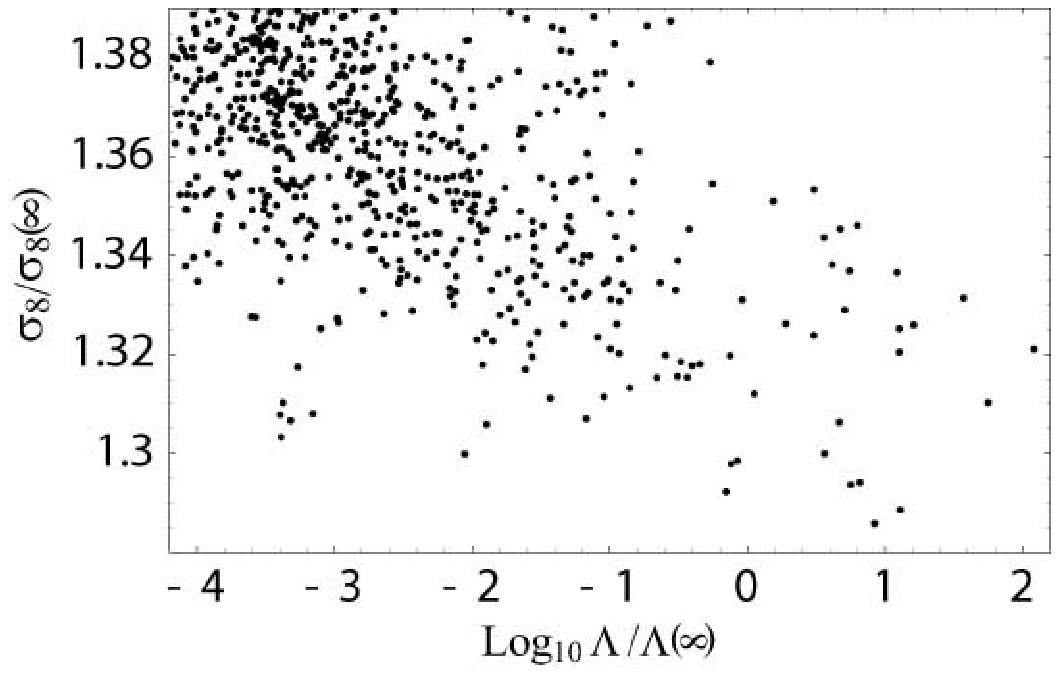,width=10.5cm}}
\caption{Plots of the 
best-fit normalization versus the 
likelihood for various orientations of the observer. 
The size of a cubic cell is $\epsilon=0.4$. 
Each quantity is normalized by the value for the simply connected 
counterpart.  }
\label{fig:8}
\EF
\pagebreak
\section{Other Topologies}
In addition to the toroidal models $T^3$ we have considered, 
there are five other flat orientable topologies, 
which can be realized as compact 
quotients of $T^3$, namely, 
$T^3/\tb{Z}_2, T^3/\tb{Z}_4,T^3/\tb{Z}_2\times \tb{Z}_2, 
T^3/\tb{Z}_3$, and $T^3/\tb{Z}_6$ where $\tb{Z}_m$ 
is a cyclic group of order $m$. The 
eigenmodes of the Laplace-Beltrami $\Delta$ operator can be
written in terms of a finite sum of eigenmodes (= plane waves)
of $T^3$, 
\BE
u_\tb{k}(\tb{x})\propto \sum_{g_i \in \Gamma} 
\exp[i \tb{k}\cdot (g_i\tb{x})], \label{eq:u}
\EE 
where $\Gamma$ denotes the discrete isometry group of $T^3$ and $\tb{x}$
is the Euclidean coordinates. 
Let us consider the case of $T^3/\tb{Z}_2$, which can be realized 
by identifying the two pairs of opposite faces of a cube with sides $L$ 
by translations but for one pair of faces with a 
translation with $\pi$ rotation. Because the finite-sheeted covering
space is described by a rectangular box with sides $L,L,2L$, one might expect
that the space can support a mode with wavelength 2$L$. However, this is
not true\cite{LSS98}. In fact, equation (\ref{eq:u}) yields the 
explicit form of the eigenmodes 
\BE
u_\tb{k}(\tb{x})\propto \exp[i \tb{k}\cdot \tb{x}]+
\exp[-i (k_1 x_1+k_2 x_2 + i k_3 (x_3+L)],
\EE  
where $k_1=2 \pi n_1/L,k_2=2 \pi n_2/L, k_3= \pi n_3/L$.
A mode with $n_1=n_2=0,n_3=1$ is equal to zero. Thus the allowed 
minimum wave number is still $k=2 \pi/L$. Similarly, one can show that 
the maximum wavelength of the eigenmodes is comparable to the
topological identification scale $L_t$ for other four 
cases\cite{LSS98}. Therefore,
it is reasonable to conclude that the constraints on these 
closed flat models do not grossly differ from those on $T^3$ models.

Although we have assumed that the background space is exactly flat so
far, it should be emphasized that a slight deviation 
from the flat space means a significant 
difference in the global topology. For instance, if the background space 
is exactly flat, there is no particular scale that sets the limit on the 
comoving size of the topological identification scale $L_t$. In other
words, one can consider arbitrarily small or large spaces. However, 
for positively curved (spherical) or negatively curved (hyperbolic) 
spaces, the curvature radius $R_c$ sets a certain 
limit on the size of the space. 

For instance, it has been proved that closed hyperbolic 
spaces with volume larger than $0.16668 \cdots 
R_c^3$ cannot exist\cite{GMT96}.
The smallest known CH space has volume  $V=0.94 R_c^3$\cite{Weeks85}. 
Remembering that the volume of a ball with radius $R$ (proper length) 
in hyperbolic space is given by $V=\pi R_c^3[\sinh (2 R/R_c)-2 R/R_c]$, the
topological identification scale of CH spaces with $V\sim R_c^3$ 
is estimated as $L_t\sim R_c$. In this case,
$\Omega_K^{-1/2}=R_c/H^{-1}\sim L_t/H^{-1}$ which gives $L_t \sim 
H^{-1}$ for $\Omega_K=0.1$ and $L_t \sim 10 H^{-1}$ for
$\Omega_K=0.01$. Thus, the effect of the global topology is
almost negligible if $\Omega_K\ll 0.1$. However, if the spatial
geometry is squashed in a certain direction while keeping the volume finite, 
the imprint of the global topology could be still prominent as in the 
case of flat topology\footnote{The fluctuations in
 ``squashed'' CH spaces resemble those in
noncompact hyperbolic spaces with cusps\cite{Inoue00b}.} although
the enhancement in the large-angle power may put a limit on the
possible shape of the fundamental cell. 
Moreover, if we allow CH orbifold models 
with a set of fixed points, then the
volume can be as small as $V=0.0391R_c^3$\cite{Inoue01b}. Then the effect of 
the nontrivial global topology is conspicuous even if the space is as
nearly flat as $\Omega_K=0.05$\cite{Inoue01b,Aurich01}. 

In the case of spherical topology, the volume is written in terms of the 
order $\vert \Gamma \vert $ of the discrete isometry group $\Gamma$
of SO$(4)$, $V=2 \pi^2 R_c^3/\vert \Gamma \vert $\cite{LL95}. 
Thus, the largest space
is the three-sphere $S^3$ itself. On the other hand, there is no lower bound 
for the volume since one can consider a group 
with arbitrarily large number of elements of $\Gamma$.  For
instance, one can consider a cyclic group $\tb{Z}_m$ with arbitrarily large
$m$. Then $V=2 \pi^2 R_c^3/\vert \tb{Z}_m \vert$ can be arbitrarily small
and the obtained quotient space is squashed in one direction. As in
CH models, the imprint of the global topology can be 
prominent for squashed spaces, even if the space is as nearly flat
as $\Omega_K=0.05$ \cite{Lehoucq02}.

\section{Conclusion}
In this paper, constraints on closed 
toroidal models from the COBE data have been reconsidered. 
By performing Bayesian analyses taking into account 
anisotropic correlation (i.e., off-diagonal elements in the covariance 
matrix) that has been often igonored in the previous literature,
we obtained a constraint on the linear size of a cubic
cell $L \gtrsim 0.6\times 2\eta_0$ for $\Omega_\Lambda\!=\!0.7,
\Omega_m\!=\!0.3$ based on the likelihoods that are marginalized over
the orientation of the observer. 
The maximum allowed number $N$ of copies of the cell
inside the last scattering surface is $\!\sim\! 2.4$ (68\% CL). We find that the
constraint becomes more stringent in comparison with that using only the
angular power spectrum. The best-fit normalization
$\sigma_8$ can be increased by $\sim$ 10 percent owing to the
suppression in the large-angle power and the globally anisotropic geometry.
%The constraint above is slightly stringent than the one in the previous
%similar analysis \cite{Inoue01a} in which the 
%fluctuations on scale $l>10$ are ignored. 
We find that the likelihood 
depends sensitively on the orientation of the observer.
Even if the likelihood that is marginalized over the orientation 
is small, the fit to the data for some limited set of orientations
is far better than that of the infinite counterpart \cite{Inoue01b}. 
If one takes the maximum likelihood value instead, the constraint becomes 
conspicuously less stringent.  
Moreover, for 
some slightly deformed toroidal models, we find that the large-angle 
power can be flat and the constraint also becomes
less stringent $N\sim 8$ (see also \cite{Roukema00}).

In order to detect the periodic structure owing to the
nontrivial topology, the topological identification scale (more
precisely twice the injectivity radius at the observing point) must be
smaller than the diameter of the last scattering surface in comoving 
coordinates: $L_t / 2 \eta_0 <1 $. Therefore, the present constraints 
do not completely deny the possibility of detection of periodic
structure by future observations\cite{CSS98a}.  

Even in the case of $L_t /2 \eta_0 \sim 1$, the signature of nontrivial
topology might be observable 
as a nontrivial correlation structure in the temperature 
correlations \cite{Levin98} because the fluctuations in the MC
models are non-Gaussian for isotropic and homogeneous observers.  
Inhomogeneous and anisotropic Gaussian fluctuations for a particular choice 
of position and orientation are regarded as non-Gaussian fluctuations
for a homogeneous and isotropic ensemble of observers\cite{FM97,Inoue00}. 
If non-Gaussian fluctuations with vanishing
skewness but non-vanishing kurtosis are found only on large angular
scales in the CMB, then it will be a strong sign of 
the nontrivial topology, or equivalently the breaking of the SCP
\cite{InoueD,KomatsuD,InoueP,Rocha}.
Even if we failed to detect the identical patterns or objects in the sky, 
the imprint of ``finiteness'' could be still observable
by measuring such statistical quantities. 
As we have discussed, the MC spaces can be squeezed irrespective of 
the type of topology. Even if the volume of the spatial section 
is large, we might be able to detect globally anisotropic structure
if the spatial section is sufficiently squeezed in a certain direction,  
The ongoing satellite missions such as MAP and 
Planck will provide us an answer to the question 
``how large is our universe?''. 
 
\section{acknowledgments}
We would like to thank K. Tomita for his continuous encouragement. 
The numerical computation was carried out on an SGI origin
3000 at Yukawa Institute Computer Facility. K.T.I. is supported 
by JSPS Research Fellowships for Young Scientists.  
This work was supported partially by Grant-in-Aid 
for Scientific Research Fund (No.14540290, No.11367).

\appendix

\section{Anisotropic Gaussian Fluctuations}

For homogeneous and isotropic Gaussian fluctuations, 
the variance of the expansion
coefficient $a_{lm}$ is constant in $m$ for a given $l$ 
regardless of the position and
orientation of the observer. In other words, there is no ``$m$-structure''
or ``direction-direction'' correlation in the expansion 
coefficients. On the other hand, for anisotropic Gaussian
fluctuations, the variance of $a_{lm}$ depends on $m$ \cite{FM97}.
 Because $a_{lm}$'s extracted from the data actually depend on $m$ 
owing to the cosmic variance, for some very 
limited choices of the orientation, the ``$m$-structure'' 
in the anisotropic fluctuations may agree with 
the apparent ``$m$-structure' in the data. 

To confirm this, we compare the variances of  
expansion coefficients $b_{lm}$ of the best-fit case to  
those extracted from the data, which can be obtained by
multiplying the temperature fluctuation $T_i$ and the standard deviation
in the noise $N(\Omega_i)$ by spherical harmonics $Y_{lm}$, and 
carrying out an integration over the pixels surviving the
``extended'' galactic cut (denoted as $S\!K$) 
\BE
b_{lm}\equiv \sum_{l'm'} a_{l'm'} W_{l'} \int_{S\!K} Y_{l'm'}(\Omega) 
Y^*_{lm}(\Omega) d \Omega + \int_{S\!K} N(\Omega) Y^*_{lm}(\Omega) d \Omega.
\EE
As shown in figure 7, an excellent agreement at $l=6-7$ is observed. 
Because the theoretical prediction of the 
quadrapole is somewhat higher than the observed
value, the large-angle suppression owing to the
finite size of the spatial section is not relevant.
We also checked other cases that give a large value in the 
likelihood compared to that of the 
simply connected counterpart and again such an excellent 
agreement in $|b_{lm}|$ at certain range of 
angular scales is observed for each case.  Thus, for a limited choice
of orientation, globally anisotropic Gaussian fluctuations can give a 
better fit compared with globally isotropic and 
homogeneous Gaussian fluctuations.

\BF
\centerline{\psfig{figure=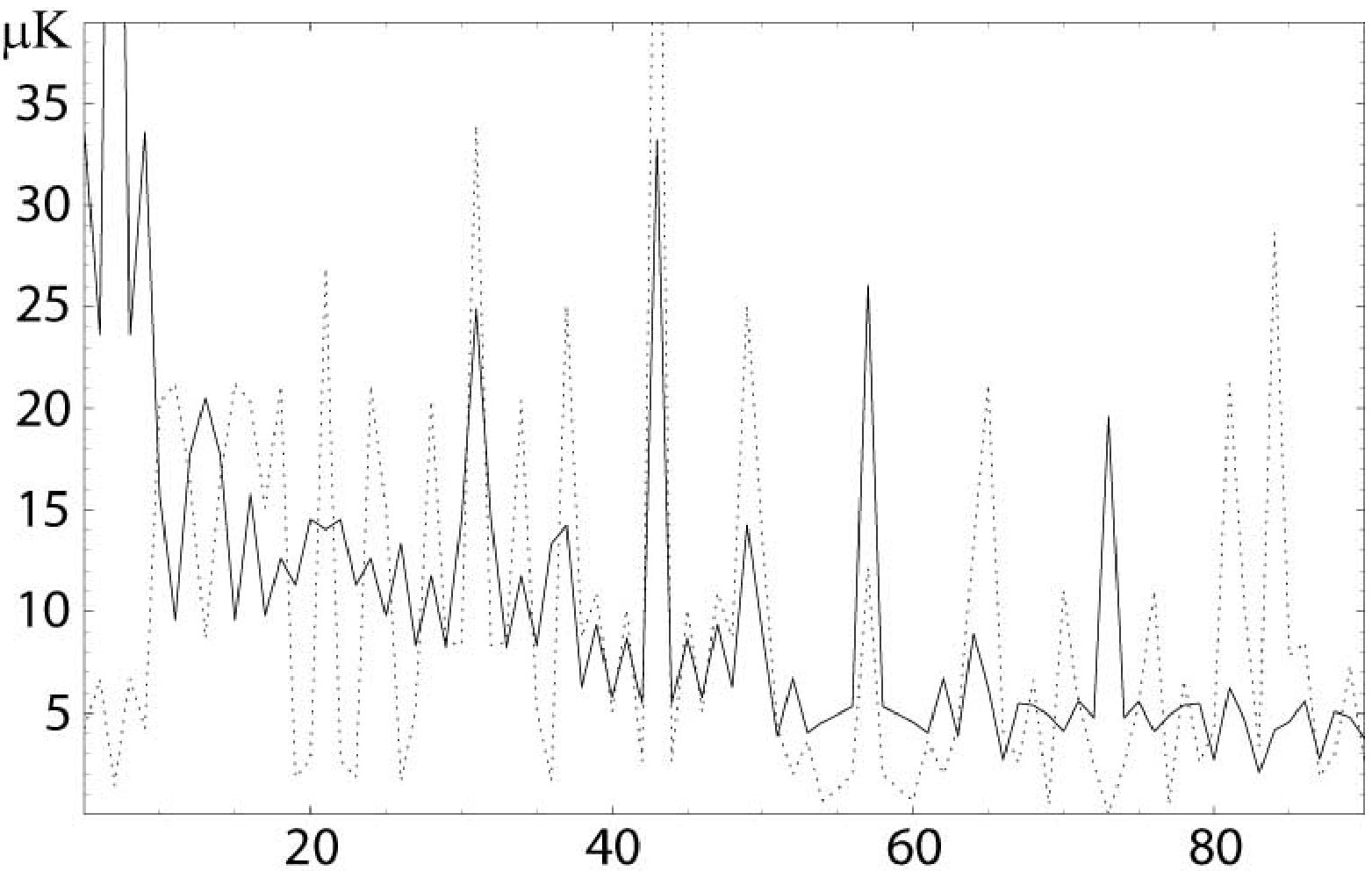,width=11cm}}
\caption{Plots of $|b_{lm}|=\sqrt{\langle b_{l m} b^*_{l m } \rangle}$ in ascending order of $l(l+1)+m+4$ for the
$T^3$ cubic model (solid curve) with $\epsilon=0.4$ (solid curve) which best fits $b_{l m}$ extracted from the COBE DMR data (dashed curve).}
\label{fig:9}
\EF

Next, we consider the effect 
of off-diagonal elements in the two-point correlation $\langle a_{lm}
a^*_{l'm'} \rangle $, $l \ne l'$ or $m \ne m'$, which represent 
scale-scale correlations.

For illustrative purpose, first, we consider a toy model in which 
the fluctuations obey a bivariate Gaussian distribution function,
\BE
p(y_1,y_2,\sigma,\tau)=
\f{1}{2 \pi \sigma^2 \sqrt{1-\tau^2}}
\exp \Biggl [
-\f{1}{2(1-\tau^2)}
\Biggr (
\f{y^2_1+y^2_2-2 \tau y_1 y_2}
{\sigma^2}
\Biggr )
\Biggr ],
\EE
where $\tau=\textrm {cov}(y_1,y_2)/\sigma^2$ is the correlation
coefficient which represents the off-diagonal components. 
By an orthogonal transformation,
$y_1=(x_1+x_2)/\sqrt{2},y_2=(-x_1+x_2)/\sqrt{2}$,
we obtain a distribution function written in terms 
of a diagonalized covariance 
matrix,
\BE
p(x_1,x_2,\sigma,\tau)=
\f{1}{2 \pi \sigma^2 \sqrt{1-\tau^2}}
\exp \Biggl [
-\f{(1-\tau)x_1^2+(1+\tau)x_2^2}
{2(1-\tau^2)\sigma^2}
\Biggr ]. \label{eq:dia}
\EE
For a given set of $x_1$ and $x_2$, 
the best-fit normalization $\sigma_{\max}$ is given by 
\BE
\sigma^2_{\max}=\f{(1-\tau)x_1^2+(1+\tau)x_2^2}{2(1-\tau^2)},
\EE
for which $p(\sigma)$ takes a maximum value. 
%\BE
%\f{d \sigma_{\max}}{d \tau}=\f{-(\tau-1)^2 x_1^2+(1+\tau)^2 x_2^2}
%{2\sqrt{2(1-\tau^2)^3} \sqrt{(1-\tau)x_1^2+(1+\tau)x_2^2)}}.
%\EE
Suppose a distribution $p$ with $\tau=0$. A slight 
increase in $\tau$ causes a squeezing of the distribution $p$
in the direction $x_1$. Then, in the region $|x_2| > |x_1|$,
$\sigma_{\max}$ increases while in the region $|x_2| < |x_1|$ 
$\sigma_{\max}$ decreases. As $\tau$ increases, the distribution is 
much squeezed and the region in which $\sigma_{\max}$ increases 
widens as $|x_2|> \sqrt{(1-\tau)/(1+\tau)}$. Therefore, in comparison
with the case in which $\tau=0$, for a wide range of parameter region
$(x_1,x_2)$, $\sigma_{\max}$ is increased if $\tau$ is large enough. 
On the other hand, the maximum value of the probability decreases, since it is 
inversely proportional to $\sigma^2_{\max}$:$p(x_1,x_2,\sigma_{\max})
=1/e\sqrt{2 \pi} \sigma^2_{\max}$.  

Although this toy model is over simplified, we can
easily apply such an argument to the two-point correlation of 
anisotropic fluctuations in the $T^3$ model.  
In that case, $y_i$ corresponds to an expansion coefficient
$a_{lm}$ and $\tau$ is generalized to $\langle a_{lm} a^*_{l'm'} \rangle/
\sqrt{C_l C_l'}$ for $l \ne \l'$ or $m \ne m'$.  

Thus, the presence of off-diagonal components in the 
covariance matrix causes a squeeze of the distribution function, 
leading to an increase in the best-fit normalization and a 
decrease in the likelihood value.

\end{document}